\documentclass[12pt,a4paper,final]{iopart}

\usepackage{amstext}
\usepackage{subfigure}
\usepackage{iopams}  
\usepackage{graphicx}
\usepackage{epstopdf}
\usepackage[breaklinks=true,colorlinks=true,linkcolor=blue,urlcolor=blue,citecolor=blue]{hyperref}

\usepackage{lineno}
\usepackage{threeparttable}
\usepackage{color}
\usepackage[section]{placeins}
\usepackage{comment}
\newcommand{\comm}[1]{}
\usepackage{epstopdf}
\usepackage{float}
\usepackage{multirow}
\usepackage{subfigure}
\usepackage{hyperref}
\usepackage{chemfig}

\modulolinenumbers[5]

\begin{document}
	
\title[\small \textit{Anisotropic-cyclicgraphene}: A new two-dimensional semiconducting carbon allotrope]{\textit{Anisotropic-cyclicgraphene}: A new two-dimensional semiconducting carbon allotrope}

\author{Marcin Ma\'zdziarz$^1$, Adam Mrozek$^2$, Wac\l{}aw Ku\'s$^3$, Tadeusz Burczy\'nski$^1$}

\address{$^1$Institute of Fundamental Technological Research Polish Academy of Sciences, Warsaw, Poland}
\address{$^2$AGH University of Science and Technology, Cracow, Poland }
\address{$^3$Silesian University of Technology, Gliwice, Poland }
\ead{mmazdz@ippt.pan.pl}

\begin{abstract}
{Potentially new, single-atom thick semiconducting 2D-\textit{graphene}-like material, called \textit{Anisotropic-cyclicgraphene}, have been generated by the two stage searching strategy linking molecular and \textit{ab initio} approach. The candidate derived from the evolutionary based algorithm and molecular simulations was then profoundly analysed using first-principles density functional theory from the structural, mechanical, phonon, and electronic properties point of view. The proposed polymorph of \textit{graphene} (\textit{rP16}-P1m1) is mechanically, dynamically, and thermally stable and can be semiconducting with a direct band gap of 0.829\,eV.}
\end{abstract}

\pacs{03.65.−w, 02.70.Ns, 73.22.−f, 62.20.−x}


\section{Introduction}
\label{sec:Int}

Carbon is probably the most versatile element of the periodic table due to the possibility of creating ({C\textbf{-}C}) bonds with different atom hybridization (\textit{sp}$^1$, \textit{sp}$^2$, and \textit{sp}$^3$) \cite{HEIMANN19971654} and its allotropes occur in all possible dimensionalities (0D-3D).
The Samara Carbon Allotrope Database (SACADA) \cite{Hoffmann2016} has now gathered more than five hundred 3D carbon allotropes.

The Nobel Prize for Physics in 2010, "for groundbreaking experiments regarding the two-dimensional material \textit{graphene}", has attracted tremendous interest in this polymorph of carbon \cite{Geim2007} due to its extraordinary physical properties \cite{Xu2014, Enyashin2011}. \textit{Graphene} is a zero-gap semiconductor with linear energy band dispersion close to the Fermi level. The absence of a band gap in the electronic spectrum of \textit{graphene} can prohibit the direct implementation of this material in electronics and that is why scientists and engineers have started to look for ways to improve this. 
It was found that mechanical deformations of \textit{graphene} can give rise to a band gap of up to 1\,eV and narrow graphene ribbons may possess the properties of semiconductors, where their band gaps are narrowed as the ribbons become wider \cite{Ivanovskii2012}.
\textit{Graphene} armchair nanotubes are metallic, whereas zigzag and chiral nanotubes can be also semiconducting with a band gap depending inversely on its diameter \cite{Lu2005}. 
Patterned defects can be utilised to disrupt the sublattice symmetry of \textit{graphene} so as to widen the band gap up to 1.2\,eV \cite{Appelhans2010}.

Undoubtedly the blossoming of studies on \textit{graphene} has accelerated interest in exploring \textit{graphene}-like materials.
Huge number of types of 2D carbon networks have been proposed, for example, by replacing some (or all)
\comm{(\ce{C=C})}(\chemfig{C=[,0.5]C}) bonds in \textit{graphene}, \textit{sp}$^2$ hybridization type, by acetylene \comm{(\ce{-C#C-})}(\textbf{\textendash}\chemfig{C~[,0.5]C}\textbf{\textendash}) linkages, \textit{sp}$^1$ hybridization type, see \cite{HEIMANN19971654, Ivanovskii2012, IVANOVSKII20131,Wang16022015,Enyashin2011}. Among these several dozens of 2D carbon allotrops, only a few are semiconductors and we will focus on those here.
Optimized geometries and electronic structures of \textit{graphyne} and its family were proposed and analysed theoretically in \cite{Baughman1987, Narita1998}. \textit{Graphyne} can be seen as a 6-fold symmetry lattice of \textit{benzene} rings connected by acetylene \comm{(\ce{-C#C-})}(\textbf{\textendash}\chemfig{C~[,0.5]C}\textbf{\textendash}) linkages. The structures proposed therein are, according to linear combination of atomic orbitals (LCAO) calculations, semiconductors with moderate band gaps. More than 30 years later, some of these structures, called \textit{graphdiyne} and \textit{Ene-yne}, were synthesized in \cite{Guoxing2010,JIA2017343}.
Semiconducting carbon allotrope named C$_z$ with a band gap of 0.858\,eV, consisting of 4, 6, 8, 12, 14 and 18 atom rings was proposed in \cite{ShuangYing2016} and similar \textit{Coro-graphene}, consisting of 4, 6 and 10 rings, with a direct band gap $\sim$0.63\,eV was found in \cite{Nulakani2015}. 
\textit{Graphenylene}, consisting of 4, 6 and 12 rings, with a direct and narrow band gap (0.025 eV) was found in \cite{Qi2013}.
\textit{T-Graphene}, 2D carbon allotrope with C$_8$ and C$_4$ rings in \cite{Liu2012} \comm{with GGA functional} is metallic, in \cite{Wang2012} as a planar structure \comm{with GGA functional} is metallic but, even as nanoribbon, is semiconducting with 0.7$\div$0.15\,eV direct band gap. However, in \cite{Nisar2012} the same planar carbon sheet can be semiconducting with a band gap, depending on the calculation method, between 0.43 and 1.01\,eV. 
A \textit{pza-C$_{10}$} allotrope consisting of alternating parallel zigzag and armchair chains (C5+C6+C7 carbon rings) was proposed in \cite{Luo2012}, where electronic structure was studied and calculated band gap was found to be 0.31 (0.71)\,eV with gradient (hybrid) functionals. 

In addition to the single-atom thick 2D structures mentioned above, multi-atom structures were also proposed.
The \textit{twin graphene} (with thickness, r$_{12}$ = 1.55\,\AA) was analysed both by first-principles methods and classical molecular dynamics simulations in \cite{Jiang2017370}. Its band gap was found to be around 1\,eV.
Theoretical two-atom thick semiconducting carbon sheet, \textit{H-net}, with indirect band gap of 0.88\,eV was postulated in \cite{Meng2014}. Even a 2D metastable carbon allotrope \textit{penta-graphene}, 2D sheet with a total thickness of 1.2{\AA} and calculated quasi-direct band gap 2.3 (4.3)\,eV with generalized gradient (Green's function) approximations was proposed in \cite{Zhang24022015,EINOLLAHZADEH20161} but questioned by \cite{Ewels22122015} because of occurrence of there tetrahedral \textit{sp}$^3$-carbon linkers. While consistent with experiment stable crystalline carbon polytypes contain only one hybridization state of carbon (either \textit{sp}$^2$ or \textit{sp}$^3$). 

The present paper is focused on an in-depth analysis of new, one-atom thick 2D carbon semiconducting material, called \textit{Anisotropic-cyclicgraphene} and is organised as outlined below: Sec.\ref{sec:Cb2Dss} references to the memetic algorithm and molecular methods used to generate \textit{Anisotropic-cyclicgraphene}, Sec.\ref{sec:Cm} briefly describes the computational \textit{ab initio} methods utilised in examination of the new structure, Sec.\ref{sec:Res} demonstrates the results of computations, and  Sec.\ref{sec:Concl} draws conclusions. 

\section{Computational methods}
\label{sec:Computationalmethods}

\subsection{Prediction of two-dimensional materials}
\label{sec:Cb2Dss}

A description and application of the memetic algorithm \cite{Mrozek2015161,Kus2016} and molecular methods \cite{Plimpton1995,Tadmor2011,Maz2011,Maz2010} using the semi-empirical potential for optimal searching for the new stable 2D \textit{graphene}-like carbon structures with predefined mechanical properties was shown in \cite{Mrozek2017}. The results obtained there, with two rectangular, primitive with 8 atoms in unit cell and P1 plane group symmetry structures:\,\textit{rP8}-P1, wherein the designations mean: 2D Pearson symbol, 2D space group, have served here as input for more accurate and reliable first-principles calculations. Consequently, the first that structure with dimensions 6.42$\times$6\,{\AA} from first-principles calculations found to be metallic,  so quite common 2D carbon material, but the second with dimensions 3.922$\times$8.472\,{\AA} found to be semiconducting, so quite rare and interesting one and only this was be analysed further. 
\subsection{\textit{Ab initio} computations}
\label{sec:Cm}

First-principles calculations with the use of density functional theory (DFT) \cite{DFT-HK, DFT-KS} within the pseudopotential, plane-wave approximation (PP-PW) have been made using the Cambridge Serial Total Energy Package (CASTEP) \cite{CASTEP}. For structural, mechanical and phonon calculations the modified Perdew-Burke-Ernzerhof generalized gradient approximation for solids (PBEsol GGA) was applied as an exchange-correlation functional \cite{Perdew2008} whereas for band structure computations the hybrid exchange-correlation functional HSE06 \cite{CASTEP_MAN,Heyd2003,Krukau2006}.
The calculation settings and methodology was taken from \cite{MAZDZIARZ20177}. 

\subsection{Finite temperature stability-molecular dynamics calculations}
\label{sec:Mdc}

The thermal stability of 2D structures is typically examined by performing \textit{ab initio} molecular dynamics (AIMD) or classical molecular dynamics (MD) with the use of interatomic potentials. Due to the significant calculation cost of AIMD,  simulations are limited to only 200 atoms and a few ps ($\sim$ 10000 steps) in \textit{NVT} (constant number of atoms, volume, and temperature) ensemble; see \cite{Nisar2012, Zhang24022015}. However to achieve reliable accuracy of the phase space sampling at least 10$^7$ steps is required \cite{Grabowski2009}, which is still too large a number for AIMD, but not for classical MD \cite{CRANFORD20114111}. 

All the molecular simulations in this work have been performed by using the Large-scale Atomic/Molecular Massively Parallel Simulator (LAMMPS) \cite{Plimpton1995} and the Adaptive Intermolecular Reactive Empirical Bound Order (AIREBO) potential for hydrocarbons \cite{Stuart2000}, and visualized through the use of the Open Visualization Tool (OVITO) \cite{Stuk2010}.

In all computations periodic boundary conditions were applied to a 2D sheet, consisting of 10$\times$10 \textit{Anisotropic-cyclicgraphene} conventional supercells in the plane of the model, and non-periodic and shrink-wrapped in the normal direction to the model \cite{Plimpton1995}, thus atoms could vibrate in 3 dimensions. Molecular dynamics simulations covered the time span of 20 ns (2x10$^7$ MD steps, one step = 1 fs). At a given temperature and zero pressure, $NPT$ (constant number of atoms, pressure and temperature) a Nose-Hoover style barostat was used \cite{Plimpton1995, Tadmor2011}. Thermodynamic information was computed and outputted every 1 ps (1000 MD steps). Similar settings were used in \cite{CRANFORD20114111} to examine thermal stability of \textit{graphyne}.

\section{Results}
\label{sec:Res}
Applying the methodology outlined in Sec.\ref{sec:Cm}, the first stage in our computations was geometric optimization of a potentially new polymorph of \textit{graphene}. 

Whereas, the initial unit cell was rectangular, primitive with 8 atoms in unit cell and P1 plane group symmetry:\,\textit{rP8}-P1, see Tab.\ref{tab:UnitCellsL} after DFT optimization it was obtained oblique, primitive with 8 atoms in unit cell and P1m1 plane group symmetry:\,\textit{oP8}-P1m1, see Tab.\ref{tab:UnitCellsB}, which corresponds to, \textit{rP16}-P1m1 conventional unit cell, see Tab.\ref{tab:UnitCells}.

\subsection{Structural Properties }
\label{ssec:RSP}
\begin{figure}[H]
	\centering
	\begin{tabular}{cc}
		\includegraphics[width=0.32\linewidth]{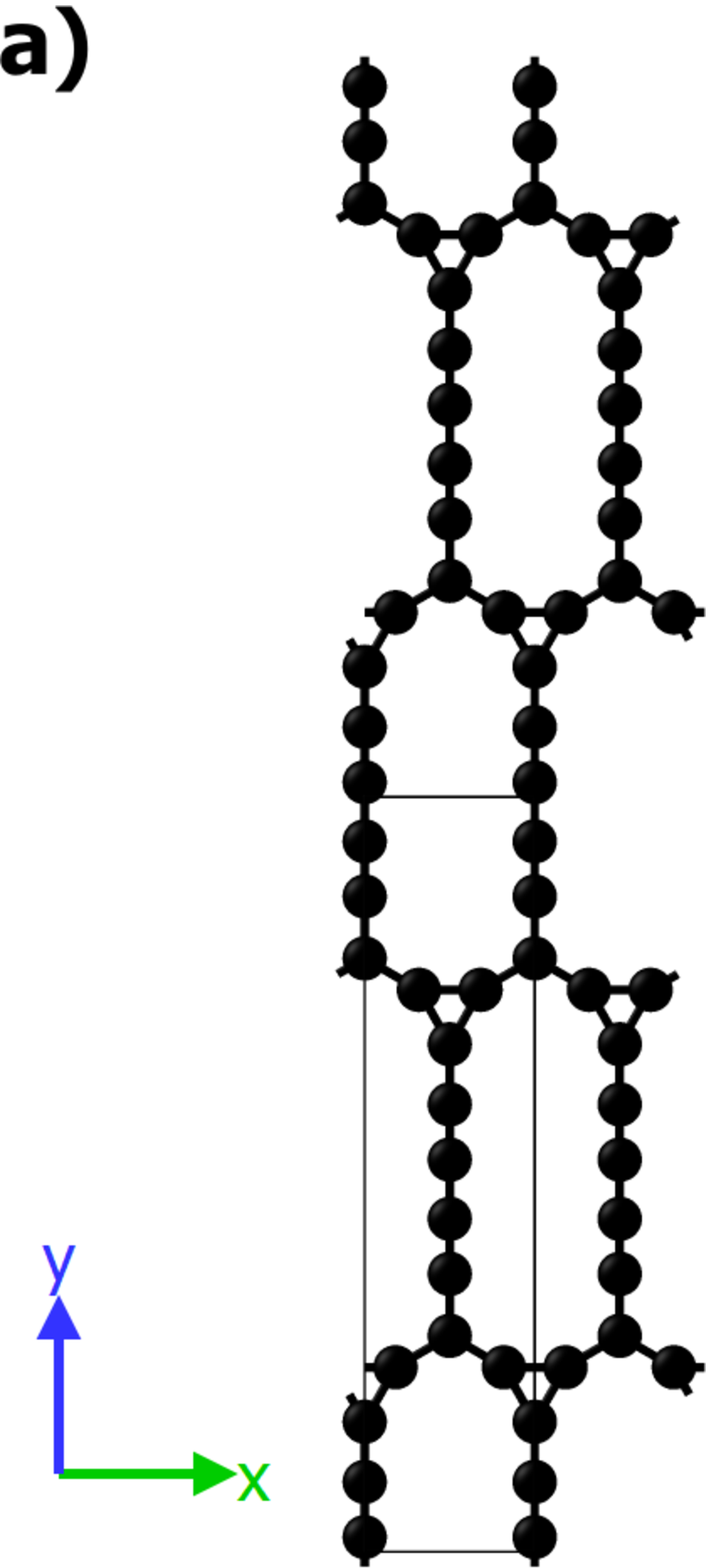} &
		\includegraphics[width=0.56\linewidth]{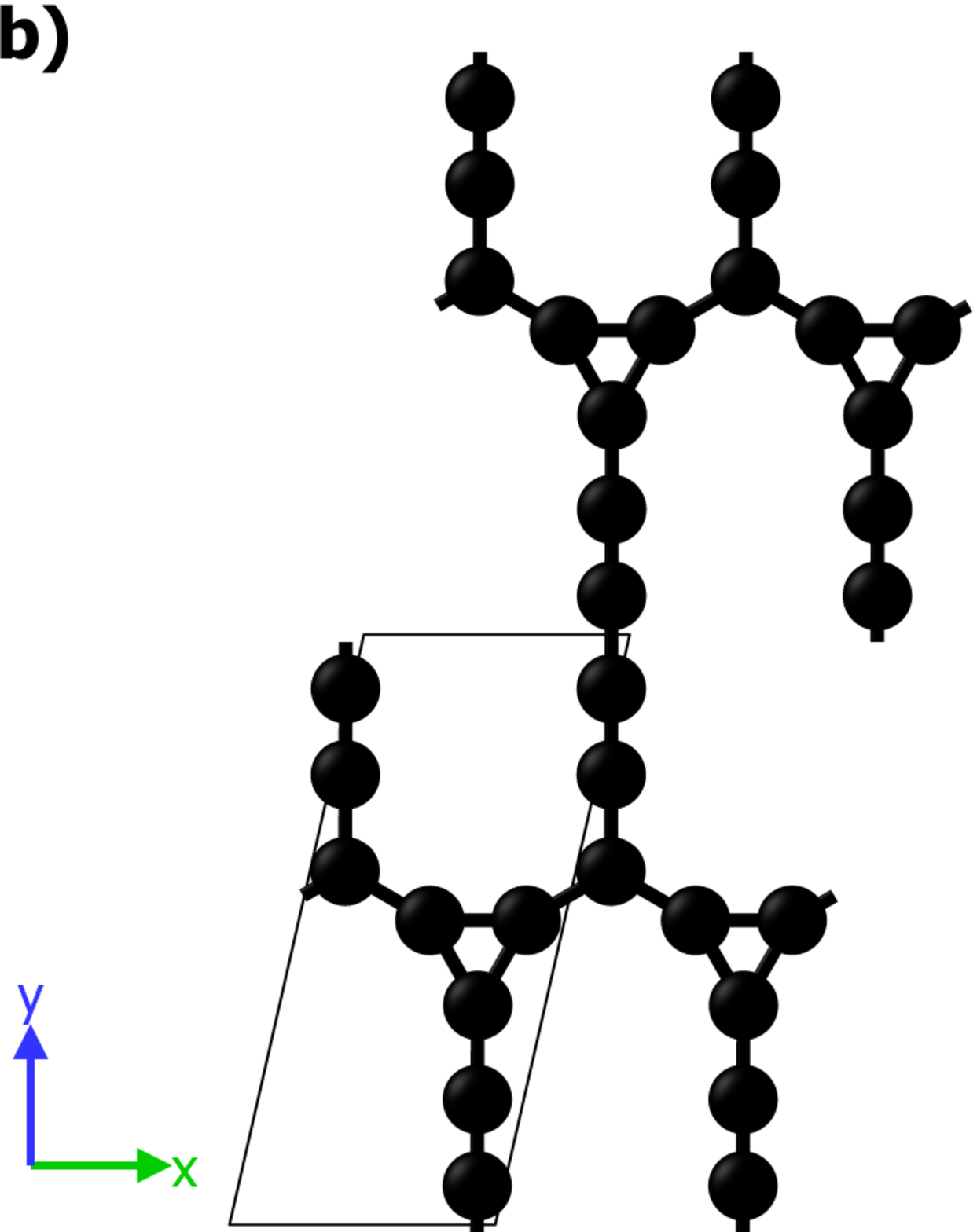} \\
	\end{tabular}
	\caption{\textit{Anisotropic-cyclicgraphene} - a)\,conventional and b)\,primitive cell  }
	\label{fig:Graphenes}
\end{figure}

For \textit{Anisotropic-cyclicgraphene}-\textit{rP16}-P1m1 (C3+C17 carbon rings), where the basic cell is portrayed in Fig.\ref{fig:Graphenes}a and recorded in Tab.\ref{tab:UnitCells}, our lattice parameters, i.e., a=3.822\,{\AA} and b=16.967\,{\AA}, are slightly smaller than those obtained from molecular calculations, see Tab.\ref{tab:OgResults}. 
It is worth mentioning that 2D non-traditional carbon materials employing a three-membered ring, i.e., C3+C12, C3+C24, C3+C36 carbon rings, were theoretically investigated in \cite{LONG20151033}, but were found to be dynamically unstable and metallic.

Unlike other single-atom thick 2D carbon semiconductors, our structure contains only such odd type of rings. 
Due to very low symmetry of the proposed structure the bond lengths vary, in the three-membered ring:  1.39-1.41\,{\AA} and in the long carbon chain: 1.24-1.39\,{\AA}. A general principle from organic chemistry says, that the more \textit{s} character the bond has, the more tightly held the bond will be and carbon bonds in representative hydrocarbons,  \chemfig{\textit{sp}^2=[,0.8]\textit{sp}^2}\,$\approx$\,1.40\,{\AA}, \chemfig{\textit{sp}-[,0.6]\textit{sp}}\,$\approx$\,1.37\,{\AA}, \chemfig{\textit{sp}^2-[,0.8]\textit{sp}^2}\,$\approx$\,1.34\,{\AA}, \chemfig{\textit{sp}~[,0.6]\textit{sp}}\,$\approx$\,1.20\,{\AA}, see \cite{fox2004organic}. Similarly here, \textit{sp}$^2$ bonds in the three-membered ring are longer than \textit{sp} bonds in long carbon chain.

If we look at the calculated cohesive energy in Tab.\ref{tab:OgResults} we see that for \textit{Anisotropic-cyclicgraphene} E$_{coh}$$\cong$-6.823\,{eV/Atom}. Cohesive energies derived from the molecular calculations are $\approx$ 0.5\,{eV/Atom} higher. The relative energy of \textit{Anisotropic-cyclicgraphene} with respect to pristine \textit{graphene} is E$_{rel}$$\cong$0.967\,{eV/Atom}, a value similar to that of other \textit{graphynes} \cite{Meng2014} and almost identical to other 2D non-traditional carbon materials employing three-membered ring as building blocks \cite{LONG20151033}, i.e., E$_{rel}$$\cong$0.96\,{eV/Atom} for the C3+C12 carbon rings and E$_{rel}$$\cong$1.03\,{eV/Atom} for C3+C24 rings and C3+C36 rings, respectively.

\subsection{Mechanical and Phonon Properties }
\label{ssec:RMP}
The in-plane elastic constants C$_{ij}$ of analysed structure are listed in Tab.\ref{tab:OgResults}. It can be seen that 
all C$_{ij}$ are lower for \textit{Anisotropic-cyclicgraphene} than for pristine \textit{graphene}, which has C$_{11}$=C$_{22}$=352.7, C$_{12}$=60.9, C$_{66}$=145.9\,(N/m) \cite{Andrew2012}. Elongated character of the basic cell, Fig.\ref{fig:Graphenes}, manifests itself in anisotropy of elastic properties. The studied structure have positive definite 2D elasticity tensor (C$_{11}$C$_{22}$ - C$^2_{12}$ $>$ 0 and C$_{66}$ $>$ 0) \cite{Zhang24022015} and is in-plane mechanically stable. The results of molecular calculations are slightly lower than those of DFT. For other \textit{graphynes} it is known that an increase in the number of acetylenic linkages corresponds to degradation of the stiffness \cite{Yue2013}.
\comm{It is known for other \textit{graphynes} that with the increase in the number of acetylenic linkages the stiffness gradually degrades \cite{Yue2013}.}

\begin{figure}[H] 
	\centering
	\begin{tabular}{cc}
		\includegraphics[width=0.44\linewidth]{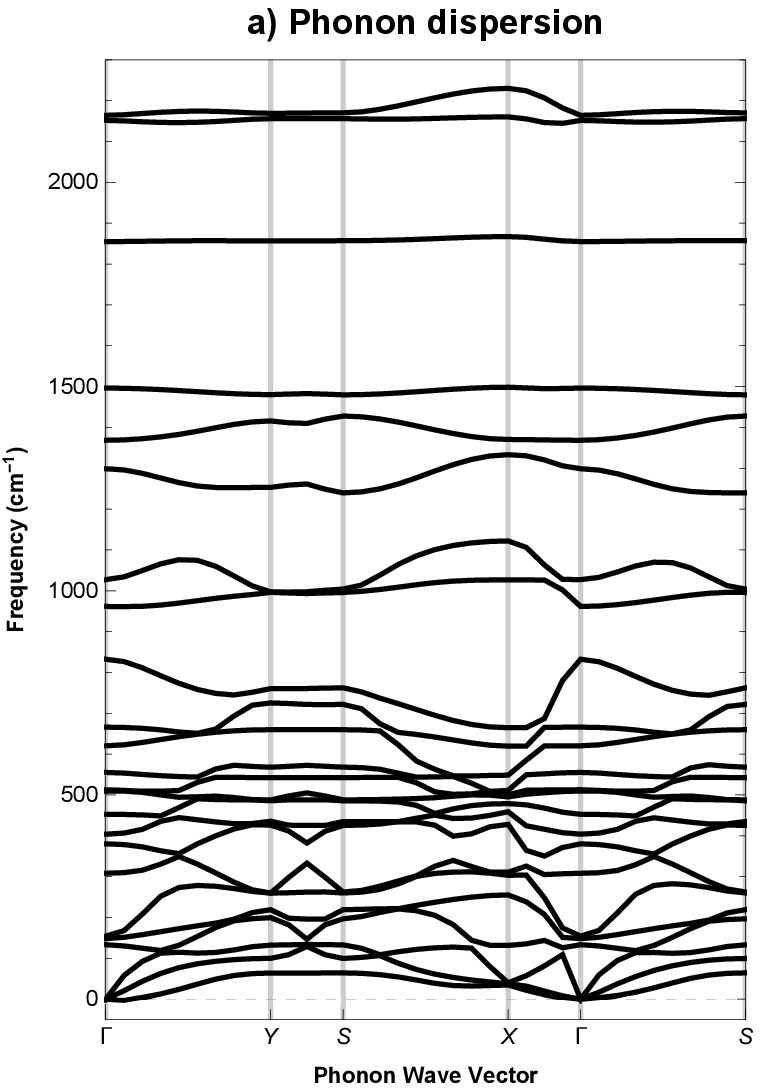} &
		\includegraphics[width=0.44\linewidth]{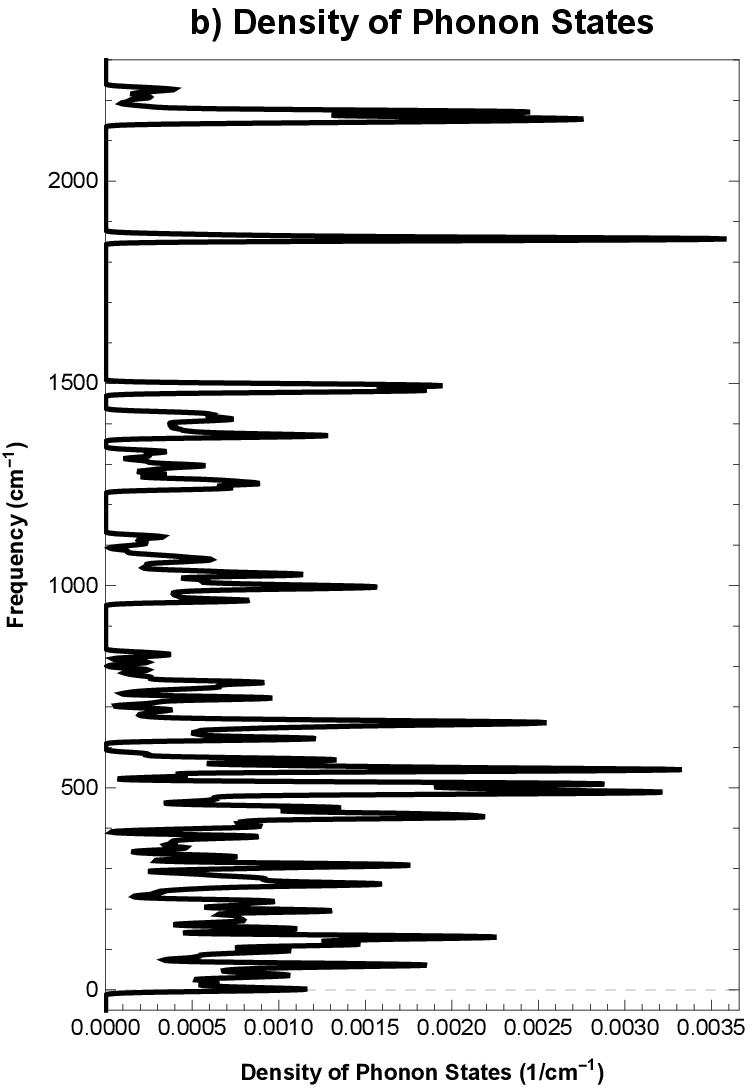}
	\end{tabular}
	\caption{\textit{Anisotropic-cyclicgraphene} - a)\,phonon dispersion and b)\,density of phonon states.}
	\label{fig:OPhononDispersion}
\end{figure}

Phonon dispersion curves for investigated structure, plotted along the high symmetry k-points, $\Gamma$(0.00, 0.00, 0.00) $\rightarrow$ Y(0.50, 0.00, 0.00) $\rightarrow$ S(0.50, 0.50, 0.00) $\rightarrow$ X(0.00, 0.50, 0.00) $\rightarrow$ $\Gamma$(0.00, 0.00, 0.00) $\rightarrow$ S(0.50, 0.50, 0.00), are drawn in Fig.\ref{fig:OPhononDispersion}. Analysis of calculated curves allows one to say that, phonon modes have positive frequencies and \textit{Anisotropic-cyclicgraphene} is not only mechanically but also dynamically stable.

\begin{table}[!htb] 
	\centering
	\centering
	\caption{Lattice parameters of conventional and primitive cell\,(\AA), cohesive energy $E_{coh}$\,(eV/Atom), relative energy $E_{rel}$\,(eV/Atom) with respect to pristine \textit{graphene}, and elastic constants $C_{ij}$\,(N/m) of \textit{Anisotropic-cyclicgraphene}.
		\label{tab:OgResults}}
	\renewcommand{\arraystretch}{1.5}
	\scriptsize 
	\begin{tabular}{|c c c|}
		\hline Source  &DFT & MD  \\
		\hline $a$ & $ 3.822 (3.822) $ & $3.936 (3.936) $ \\
		$b$ & $16.967 (8.701)$ & $17.544 (9.000)$ \\
		$\gamma$ & $90.0^{\circ} (77.166^{\circ}) $ &  $90.0^{\circ} (77.17^{\circ})$ \\
		$E_{coh}$ & $-6.823 $ & $-6.292$ \\
		$E_{rel}$ & $0.967 $ & $1.133 $ \\
		$C_{11}$ & $80.54$ & $71.17 $ \\
		$C_{22}$ & $294.46$ & $257.45 $ \\
		$C_{12}$ & $47.19$ & $46.41 $ \\
		$C_{66}$ & $5.38$ & $2.79 $ \\
		\hline 
	\end{tabular}
\end{table}

\subsection{Electronic Properties }
\label{ssec:REP}

\begin{figure}[H] 
	\centering
	\begin{tabular}{cc}
		\includegraphics[width=0.44\linewidth]{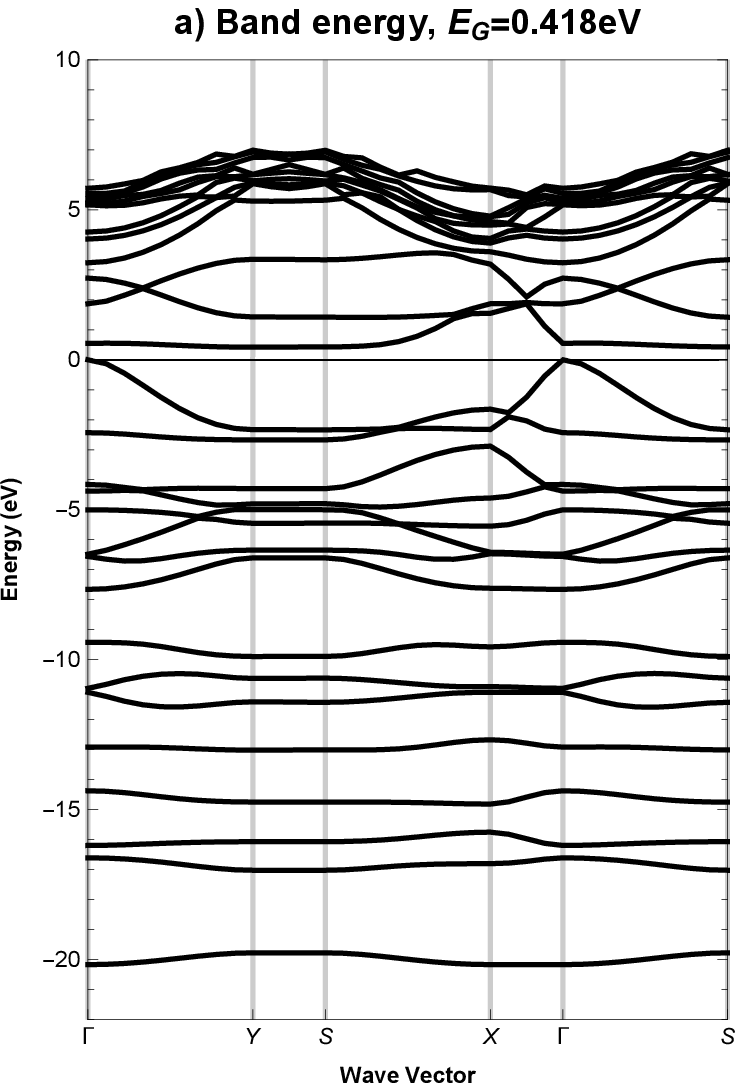} &		
		\includegraphics[width=0.44\linewidth]{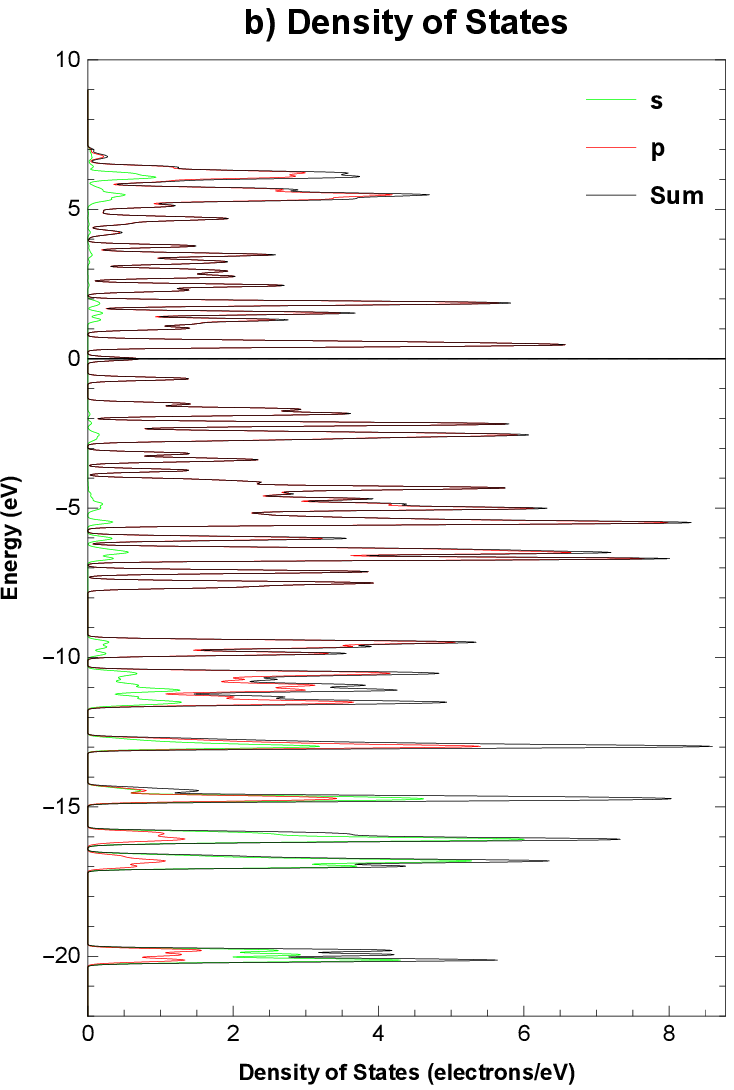}
	\end{tabular}
	\caption{\textit{Anisotropic-cyclicgraphene} - a)\,electronic band structure and b)\,total and partial density of states\,(GGA functional). The Fermi level E$_{F}$ is set to 0.}
	\label{fig:OBandenergy}
\end{figure}

\begin{figure}[H] 
	\centering
	\begin{tabular}{cc}
		\includegraphics[width=0.44\linewidth]{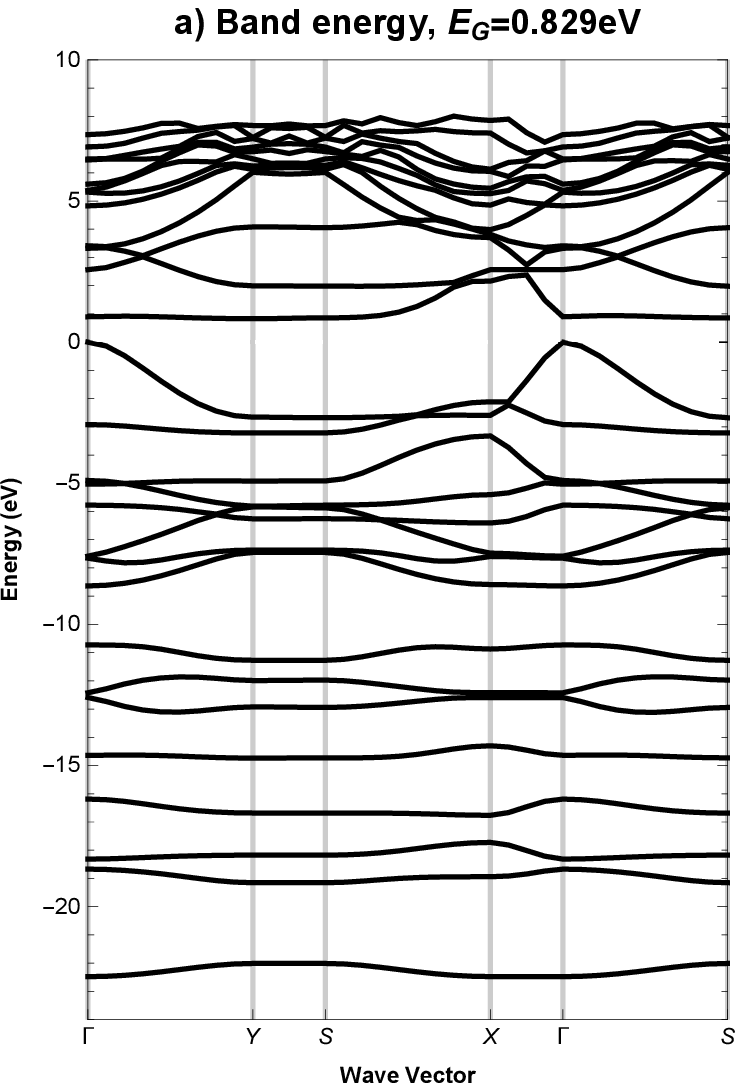} &
		\includegraphics[width=0.44\linewidth]{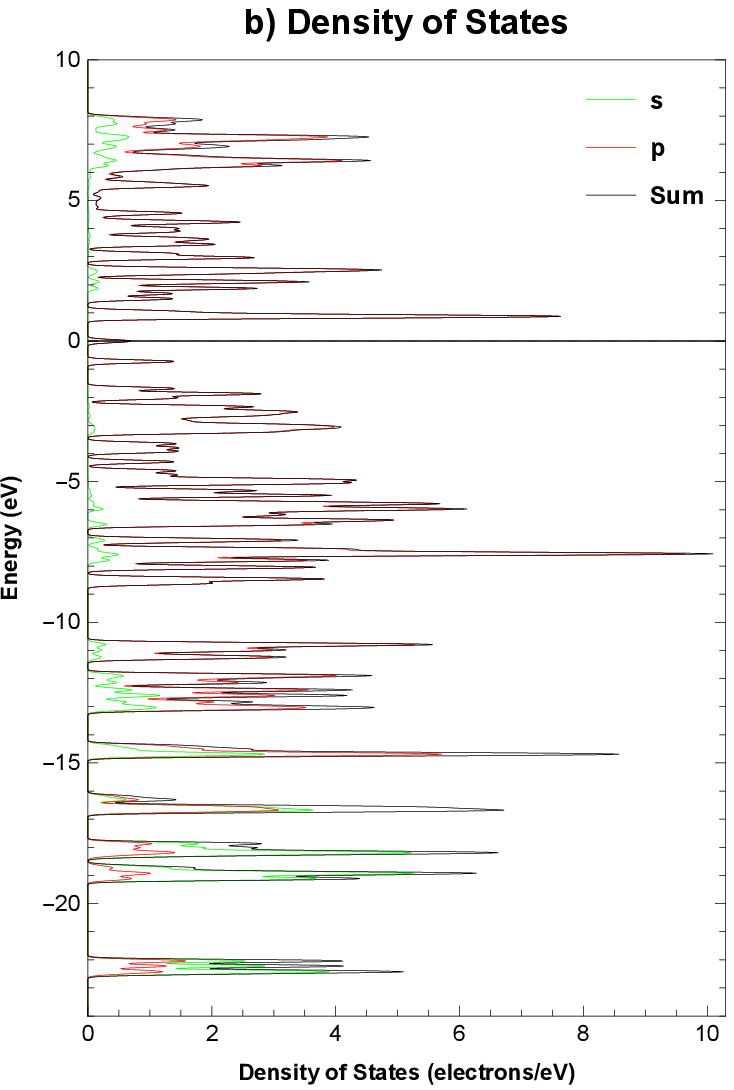} 
	\end{tabular}
	\caption{\textit{Anisotropic-cyclicgraphene} - a)\,electronic band structure and b)\,total and partial density of states\,(HSE06 functional). The Fermi level E$_{F}$ is set to 0.}
	\label{fig:OBandenergyH}
\end{figure}

\begin{figure}[H] 
	\centering
	\begin{tabular}{ccc}
		\includegraphics[width=0.44\linewidth]{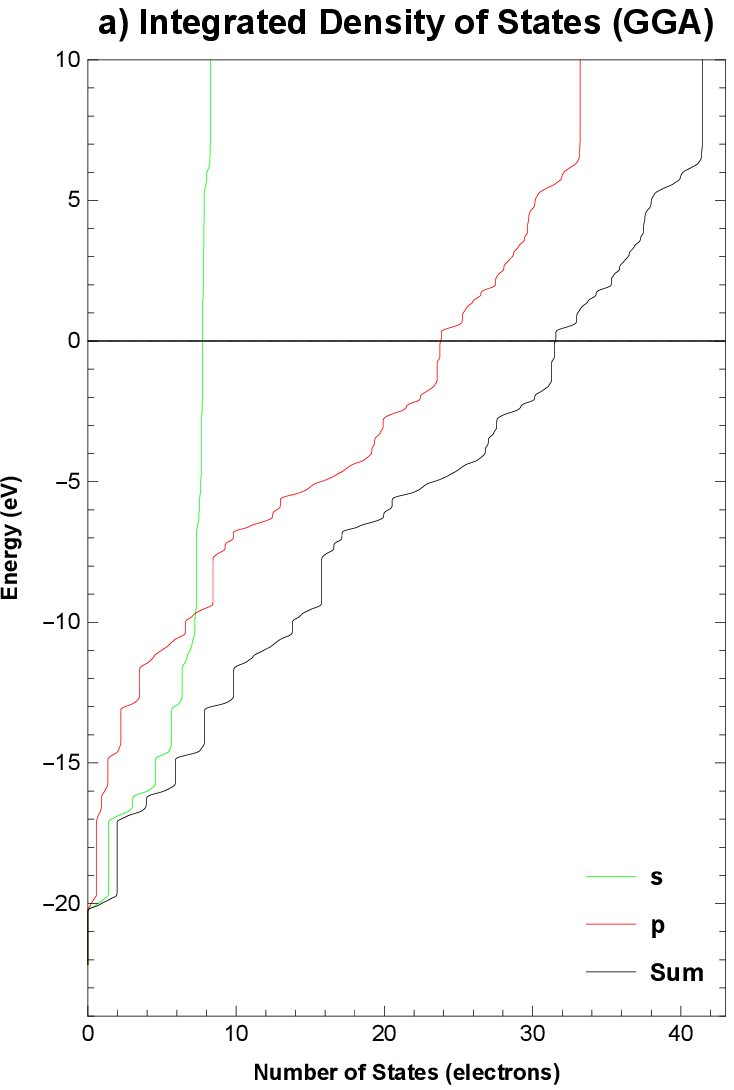} &
		\includegraphics[width=0.445\linewidth]{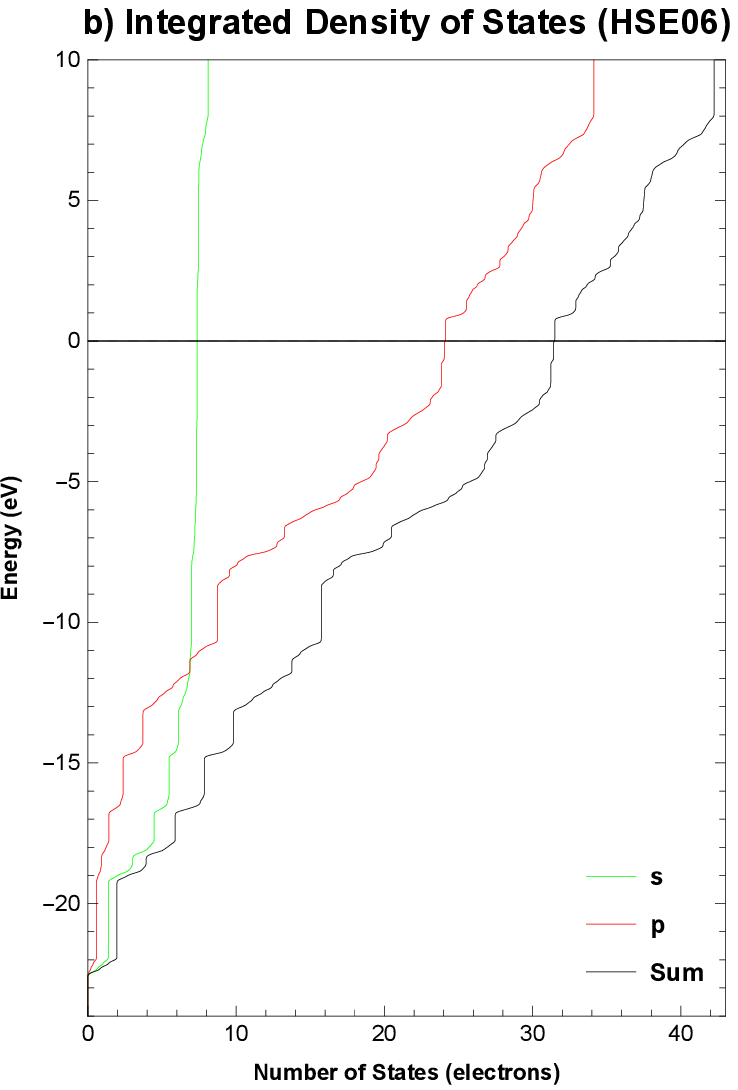} 
	\end{tabular}
	\caption{\textit{Anisotropic-cyclicgraphene} - integrated total and partial density of states - a)\,(GGA functional) and b)\,(HSE06 functional). The Fermi level E$_{F}$ is set to 0.}
	\label{fig:IOBandenergy}
\end{figure}

The electronic band structure (EBS) and density of states (DOS), total and partial, for the analysed structure are plotted along the high symmetry \textit{k}-points, $\Gamma$(0.00, 0.00, 0.00) $\rightarrow$ Y(0.50, 0.00, 0.00) $\rightarrow$ S(0.50, 0.50, 0.00) $\rightarrow$ X(0.00, 0.50, 0.00) $\rightarrow$ $\Gamma$(0.00, 0.00, 0.00) $\rightarrow$ S(0.50, 0.50, 0.00), in Fig.\ref{fig:OBandenergy} for the GGA functional and in Fig.\ref{fig:OBandenergyH} for the HSE06 hybrid functional. The integrated total and partial density of states for both functionals are presented in Figs.\ref{fig:IOBandenergy}. 
The overall pattern of the valence band (VB) and conduction band (CB) of the analysed structure is quite similar for both functionals, however, it should be noted, that for HSE06 hybrid functional the VB is a bit wider and the CB is wider and shifted slightly upwards than for GGA functional, see Fig.\ref{fig:OBandenergy}\,b) and Fig.\ref{fig:OBandenergyH}\,b). A direct band gap is E$_g$=0.418\,eV for the functional PBEsol, while E$_g$=0.829\,eV for the HSE06 at the $\Gamma$ point of the first Brillouin zone. Analysis of integrated total and partial density of states in Figs.\ref{fig:IOBandenergy} shows that from the energy level of approximately -11\,eV, the \textit{s}-states minimally contribute to the total DOS and it is clearly evident that the \textit{p}-states are primarily responsible for the formation of VB and CB near to the Fermi level.
The calculated Fermi energy for \textit{Anisotropic-cyclicgraphene} system equals -4.243\,eV with the PBEsol GGA functional and -3.106\,eV with the HSE06 hybrid functional.

\subsection{Finite temperature stability}
\label{sec:RFts}

\begin{figure}[H] 
	\centering
	\begin{tabular}{cc}
		\includegraphics[width=0.44\linewidth]{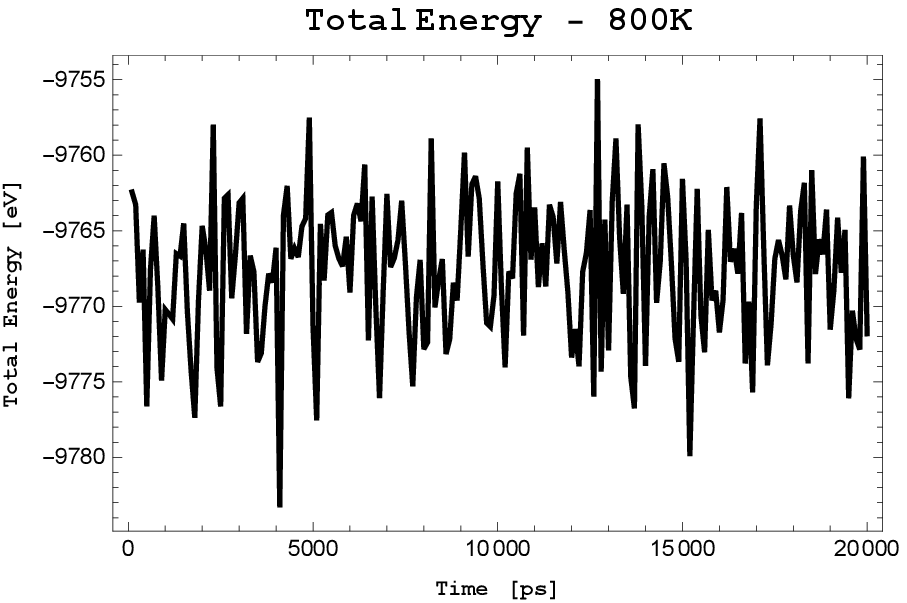} &
		\includegraphics[width=0.44\linewidth]{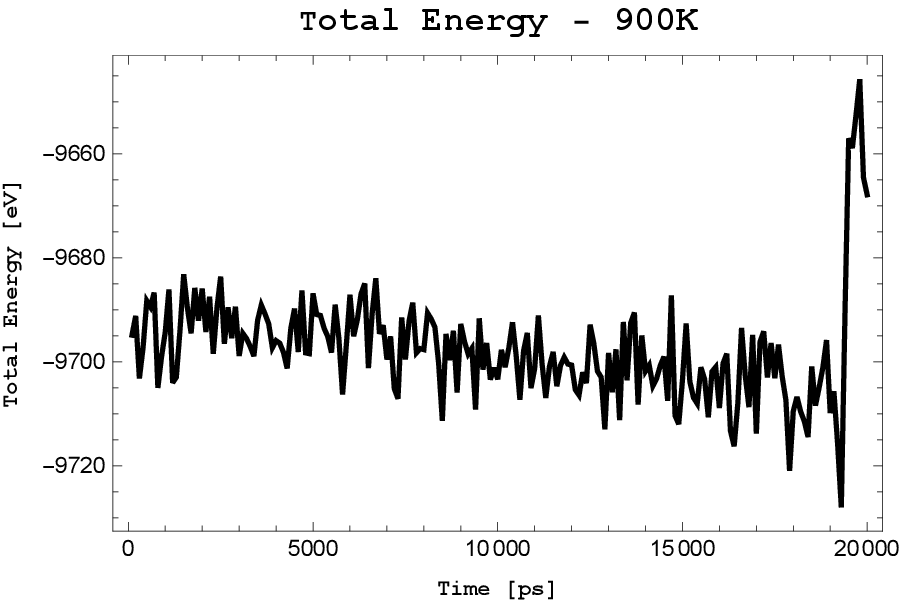} 
	\end{tabular}
	\caption{\textit{Anisotropic-cyclicgraphene} - total energy for 800\,K and 900\,K }
	\label{fig:TE}
\end{figure}

\begin{figure}[H] 
	\centering
	\begin{tabular}{cc}
		\includegraphics[width=0.44\linewidth]{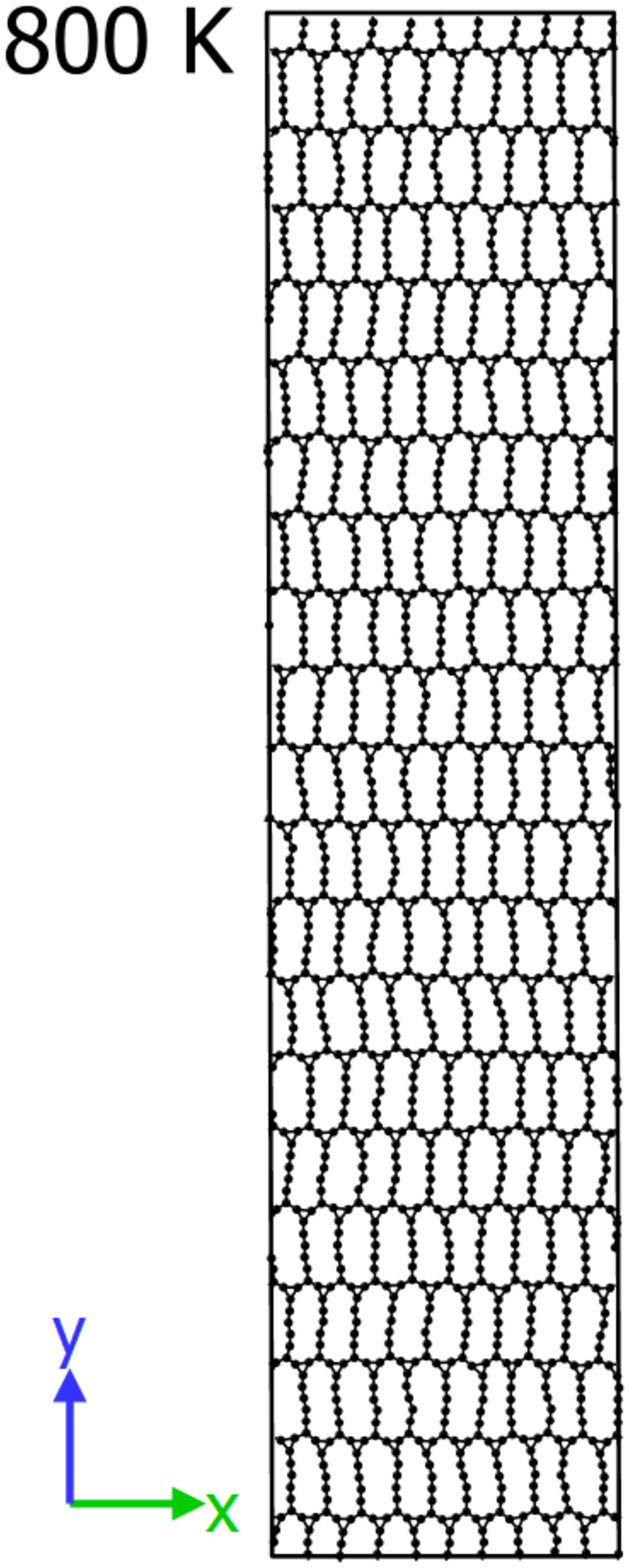} &
		\includegraphics[width=0.44\linewidth]{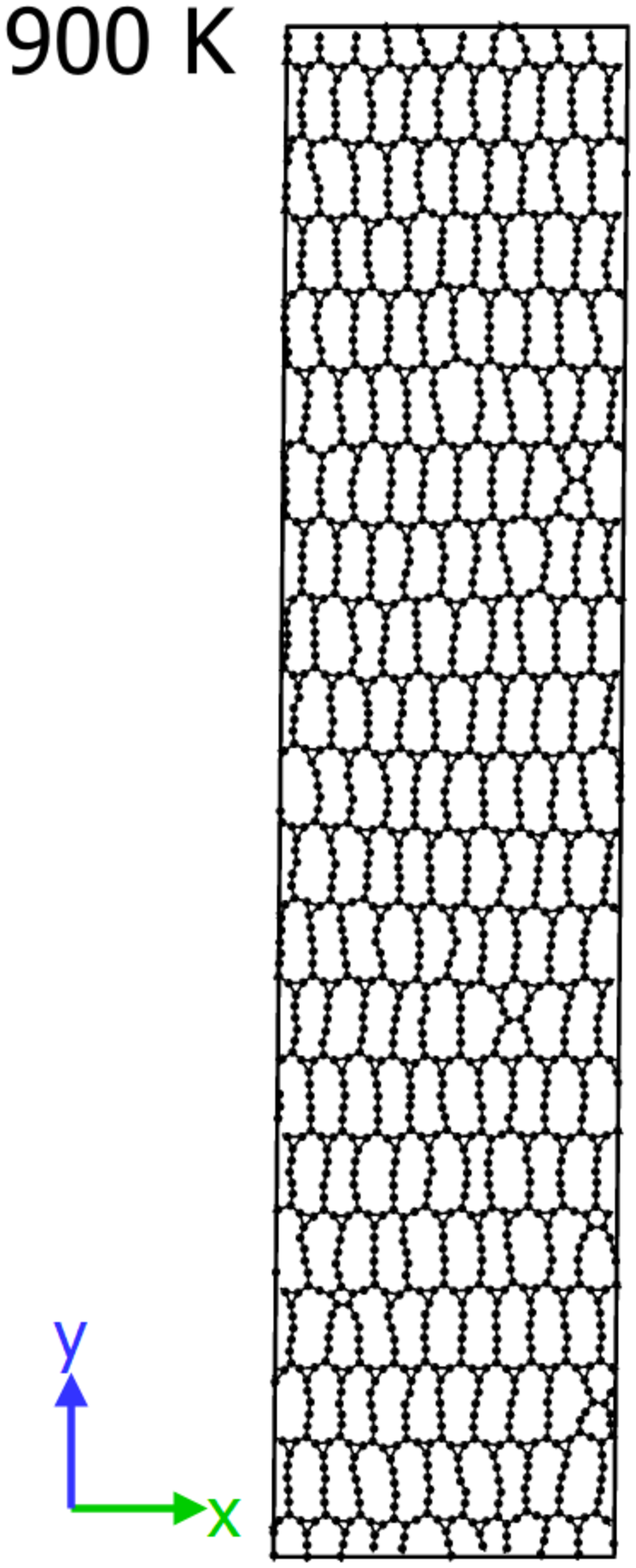}
	\end{tabular}
	\caption{\textit{Anisotropic-cyclicgraphene} - snapshots of the structures at 800\,K and 900\,K after 20\,ns}
	\label{fig:MDsnap}
\end{figure}

We see in Fig.\ref{fig:TE} that the total energy for T=800\,K fluctuate around average value throughout the simulation, whereas for T=900\,K it decreases slowly until a considerable change at a time $\sim$18\,ns. By analysing snapshots of the structures at given temperature  in  Fig.\ref{fig:MDsnap}, we see that for T=800\,K the topology of \textit{Anisotropic-cyclicgraphene} does not change and atoms vibrate around equilibrium positions. At T=900\,K topological changes occur and many smaller rings appear. We therefore conclude that the polymorph is thermally stable up to a temperature of 800\,K.

\section{Conclusions}
\label{sec:Concl}

The extensive analysis of potentially new polymorph of \textit{graphene} consisting of C3+C17 carbon anisotropic rings within the framework of DFT from point of view of the structural, mechanical, phonon, and electronic properties was carried out in the paper. All above calculations have been completed using \comm{were realized with} ultra-fine quality settings, the modified PBEsol GGA for solids exchange-correlation functional and additionally for electronic band structure computations the hybrid exchange-correlation functional HSE06. In addition the thermal stability of the proposed \textit{Anisotropic-cyclicgraphene} was examined by performing series of classical molecular dynamics simulations.

The following conclusions can be stated:
\begin{itemize}
	\item The proposed polymorph of \textit{graphene} (\textit{rP16}-P1m1) is mechanically and dynamically stable contrary to other C3-\textit{cyclicgraphenes}.  
	\item The proposed structure is thermally stable up to a temperature of at least 800\,K.
	\item The relative energy of \textit{Anisotropic-cyclicgraphene} \comm{\textit{O-graphene}} with respect to pristine \textit{graphene} is similar to other \textit{graphynes} and \textit{cyclicgraphenes}.  
	\item \textit{Anisotropic-cyclicgraphene} can be semiconducting, with a direct band gap with a value of 0.829\,eV.
	\item The semi-empirical potential AIREBO seems to be surprisingly suitable for carbon structures.
\end{itemize}

Some results in this paper are unique and that we trust will be verified by other works. Naturally, the synthesis of the proposed structure is a separate task and goes beyond the area of this work. The effect of carbon rings anisotropy and chain lengths on the properties of potentially new 2D structures will be the subject of further research.

\section*{ACKNOWLEDGMENTS}
This work was partially supported by the National Science Centre (NCN -- Poland) Research Project: UMO-2016/21/B/ST8/02450. Access to the program CASTEP was provided by the Interdisciplinary Centre for Mathematical and Computational Modelling of Warsaw University (ICM UW). 

\appendix 
\section{Unit Cells}

The following tables present crystallographic data for \textit{Anisotropic-cyclicgraphene} -\textit{rP16}-P1m1.

\label{sec:App}
\begin{table}[H] 
	\centering
	\centering
	\caption{\label{tab:UnitCells}Lattice parameters\,(\AA) and fractional coordinates of atoms of \textit{Anisotropic-cyclicgraphene} - \textit{rP16}-P1m1-conventional cell, Fig.\ref{fig:Graphenes}a.
		}
	\renewcommand{\arraystretch}{1.5}
	\tiny
	\begin{tabular}{|c c c|}
		\hline \multicolumn{3}{ |c| } {Lattice parameters} \\
		\hline $a$ & \multicolumn{2}{ c| }{3.822}  \\
		$b$ & \multicolumn{2}{ c|}{16.967} \\
		\hline {  Atom Number} & \multicolumn{2}{ c| } {Fractional coordinates of atoms} \\
		\hline 1 & 0.50000 & 0.28603 \\
		 2 & 0.50000 & 0.36811 \\
		 3 & 0.50000 & 0.44107 \\
	     4 & 0.00000 & 0.78603 \\
	     5 & 0.00000 & 0.86811 \\
	     6 & 0.00000 & 0.94107 \\
	     7 & 0.18150 & 0.24436 \\
	     8 & 0.68150 & 0.74436 \\
	     9 & 0.81850 & 0.24436 \\
	    10 & 0.31850 & 0.74436 \\ 
	    11 & 0.00000 & 0.17228 \\
	    12 & 0.00000 & 0.01941 \\
	    13 & 0.00000 & 0.09259 \\
	    14 & 0.50000 & 0.67228 \\
	    15 & 0.50000 & 0.51941 \\
	    16 & 0.50000 & 0.59259 \\
		\hline 
	\end{tabular}
\end{table}

\begin{table}[H] 
	\centering
	\centering
	\caption{\label{tab:UnitCellsB}Lattice parameters\,(\AA) and fractional coordinates of atoms of \textit{Anisotropic-cyclicgraphene} - \textit{oP8}-P1m1-primitive cell, Fig.\ref{fig:Graphenes}b.
	}
	\renewcommand{\arraystretch}{1.5}
	\tiny
	\begin{tabular}{|c c c|}
		\hline \multicolumn{3}{ |c| } {Lattice parameters} \\
		\hline $a$ & \multicolumn{2}{ c| }{3.822}  \\
		$b$ & \multicolumn{2}{ c|}{8.701} \\
		$\gamma$ & \multicolumn{2}{ c|}{77.1665$^{\circ}$} \\
		\hline {  Atom Number} & \multicolumn{2}{ c| } {Fractional coordinates of atoms} \\
		\hline 1 & 0.13221 & 0.59914 \\
		2 & 0.85619 & 0.51591 \\
		3 & 0.49331 & 0.51566 \\
		4 & 0.05066 & 0.76329 \\
		5 & 0.97724 & 0.90920 \\
		6 & 0.74632 & 0.37162 \\
		7 & 0.89892 & 0.06589 \\
		8 & 0.82547 & 0.21225 \\
		\hline 
	\end{tabular}
\end{table}
 
 \begin{table}[H] 
 	\centering
 	\centering
 	\caption{\label{tab:UnitCellsL}Lattice parameters\,(\AA) and fractional coordinates of atoms of initial \textit{cyclicgraphene} - \textit{rP8}-P1-primitive cell, \cite{Mrozek2017}.
 	}
 	\renewcommand{\arraystretch}{1.5}
 	\tiny
 	\begin{tabular}{|c c c|}
 		\hline \multicolumn{3}{ |c| } {Lattice parameters} \\
 		\hline $a$ & \multicolumn{2}{ c| }{3.922}  \\
 		$b$ & \multicolumn{2}{ c|}{8.472} \\
 		$\gamma$ & \multicolumn{2}{ c|}{90.0$^{\circ}$} \\
 		\hline {  Atom Number} & \multicolumn{2}{ c| } {Fractional coordinates of atoms} \\
 		\hline 1 & 0.15902  &   0.60266 \\
 		2 & 0.85954   &   0.52400 \\
 		3 & 0.45448   &   0.52020 \\
 		4 & 0.14158   &   0.76639 \\
 		5 & 0.02079   &   0.91373 \\
 		6 & 0.67188   &   0.35866 \\
 		7 & 0.89386   &   0.05964 \\
 		8 & 0.77916   &   0.20769 \\
 		\hline 
 	\end{tabular}
 \end{table}
 










\bibliographystyle{unsrt} 
\section*{References}
\bibliography{References}
\end{document}